\documentclass{article}

\usepackage{arxiv}

\usepackage[utf8]{inputenc} 
\usepackage[T1]{fontenc}    
\usepackage{hyperref}       
\usepackage{url}            
\usepackage{booktabs}       
\usepackage{amsfonts}       
\usepackage{nicefrac}       
\usepackage{microtype}      
\usepackage{lipsum}		
\usepackage{graphicx}
\usepackage{natbib}
\usepackage{doi}

\usepackage{subfig}
\usepackage{multirow}
\usepackage[separate-uncertainty=true]{siunitx}

\title{The Construction of a Soft Gripper\\Based on Magnetorheological Elastomer with Permanent Magnet}

\author{
  Jakub Bernat, Pawel Czopek, Paulina Superczynska \\
  Institute of Automatic Control and Robotics \\
  Poznan University of Technology \\
  Poznan, Poland\\
  \texttt{\{jakub.bernat, paulina.superczynska\}@put.poznan.pl} \\
   \And
  Piotr Gajewski, Agnieszka Marcinkowska \\
  Institute of Chemical Technology and Engineering \\
  Poznan University of Technology \\
  Poznan, Poland\\
  \texttt{\{piotr.gajewski, agnieszka.marcinkowska\}@put.poznan.pl} \\
}

\renewcommand{\shorttitle}{The Construction of a Soft Gripper Based on MRE with Permanent Magnet}

\hypersetup{
pdftitle={The Construction of a Soft Gripper Based on MRE  with Permanent Magnet},
pdfsubject={q-bio.NC, q-bio.QM},
pdfauthor={Jakub Bernat, Pawel czopek, Paulina Superczynska, Piotr Gajewski, Agnieszka Marcinkowska},
pdfkeywords={magnetorheological elastomer, gripper, permanent magnet, rolling effect},
}

\begin{document}
\maketitle

\begin{abstract}
Recently, magnetorheological elastomers have become an interesting smart material with many new designs for robotics. A variety of applications have been built with magnetorheological elastomers, such as vibration absorbers, actuators, or grippers, showing that this material is promising for soft robotics. In this work, the novel concept of a gripper is proposed, exploring the features of a magnetorheological elastomer and permanent magnet. The gripper uses the energy of a permanent magnet to provide a self-closing gripping mechanism. The usage of flexible material enables one to hold delicate objects of various shapes. This paper presents the rolling effect of magnetorheological elastomer and permanent magnet, the design process, and the features of the soft gripper. The effectiveness of the soft gripper was validated in a series of experiments that involved lifting different objects.
\end{abstract}

\section{Introduction}

In recent times, soft magnetic materials have become more important in robotics and control systems applications \cite{rateni2015design,erb2016magnetic,Sitti-RSS-19,ESHAGHI2021101268}. The classic usage of soft magnetic materials is mainly related to vibration control \cite{li2014state,Lin2023109633}. Its flexibility and ability to change the mechanical properties under the control of a magnetic field allow the construction of vibration isolators and absorbers in commercial applications. However, recently, these materials became more interesting for robotic applications \cite{erb2016magnetic,bira2020review,bose2021magnetorheological}, including the construction of haptic devices, soft grippers, and remote control of the robot by an external magnetic field\cite{Mahoney2014,Popek-Magnet7989138,Culha2020}. 

In the literature, two kinds of magnetoactive material have been described. The first is a magnetorheological elastomer (MRE), an elastomer mixed with ferromagnetic particles. These materials' preparation process and properties can be found in detail in the work \cite{li2014state,bira2020review,bose2021magnetorheological,bernat2022gripper}. The main properties of MRE are flexibility and reaction to the magnetic field due to a relative permeability of about 2-5. The cost of MRE material flexibility is lower permeability in comparison to rigid materials such as steel. The second magnetoactive material is an elastomer mixed with a hard magnetic material. It is also flexible, but it has a magnetic field. The fabrication process of this material can be seen in the work \cite{erb2016magnetic}. In this work, we focus on the first type of magnetorheological elastomer in robotics applications. The main problem with MRE material in robotic applications is its low permeability, which limits the design possibilities. As previous studies show, MRE requires strong magnetic fields \cite{bose2021magnetorheological,Cramer2018127} to be controlled or to produce large deformations. In the work \cite{Cramer2018127} it is suggested that MRE requires further improvement of material parameters before being used in robotic applications. However, in our work, we show that MRE is already applicable to creating soft grippers.

In this work, we are particularly interested in the construction of the soft gripper, which in recent times has been strongly developed by researchers \cite{Rus2015ER,rateni2015design,navas2021soft,shintake2018soft,Pagoli2022}. One of the applications in which soft grippers can be used in the food industry \cite{pettersson2010design, dinakaran2023}, because of the possibility to grasp and transport delicate products such as fruits or candies without damaging them. In general, soft grippers can be divided by the actuation principle. The most popular soft grippers are pneumatic and mechanical actuation. The primary role of soft pneumatic grippers represents PneuNet \cite{mosadegh2014pneumatic} or particle jamming \cite{Rus2015ER,Washio9982126}. They are controlled by variable pressure and therefore must be supplied with compressed air. On the other hand, the soft gripper's fingers are moved by a cable-based mechanism \cite{rateni2015design}, which represents a mechanical type. The soft gripper with pneumatic and mechanics actuation has good properties, however, they require a completed control system like an air compressor. Therefore, grippers based on actuation principles different from pneumatic or mechanical ones also play an important role. For example, in works \cite{Araromi:6844878,Shintake:7353507} the dielectric electroactive polymers are applied to build grippers. 

An interesting alternative to building soft grippers is the application of magnetoactive materials with soft or hard magnetic particles. In the literature \cite{Zhang20218181,Skfivan:8722762,Choi202044147,bernat2022gripper,Guan2022}, there exist some preliminary studies on how to build a gripper using MRE material (with soft magnetic particles). In the work \cite{Skfivan:8722762} the MRE fingers, which are based on a rigid skeleton, are controlled by varying magnetic fields. The gripper can hold varying objects and requires energy in the closed state. In work \cite{Zhang20218181}, the soft gripper is performed on an electromagnet and MRE membrane to create a suction cup. The studies \cite{Choi202044147} show a mechanical gripper where the adaptive sking is built with MRE material. The gripper can adapt to different object shapes hence it can hold objects more delicate and precise. In the work \cite{bernat2022gripper,Guan2022} the initial concept of the MRE gripper with fingers was designed using a magnetorheological material controlled by an electromagnet. This gripper holds only lightweight objects and its gripping force is created only by stress caused by deformation from the held object. Finally, MRE can also be an auxiliary material for building a gripper like in works \cite{Choi2018,Choi2023}, where the MRE creates a bladder for magnetorheological fluid. Another application of MRE to grippers as auxiliary material is its connection to shape memory alloy (SMA) presented in \cite{Yang20224585}. It is only responsible for improving the performance of SMA. As an alternative to MRE, the application of hard magnetic particles with flexible polymer is also used to build soft grippers, as shown in the work \cite{ullrich2015magnetically,pettersson2010design,Ding2023,Gao2022}. In this topic, the interesting review of magnetoactive materials with hard magnetic particles for the construction of microgrippers is presented in the work \cite{ESHAGHI2021101268} where the most common configuration is to control the gripper by varying external fields. It is worth noticing that magnetoactive materials with hard magnetic particles are more difficult to fabricate than MRE.

The main goal of the presented work is to investigate a novel geometry concept for the MRE gripper. It is based on the interaction between the permanent magnet and the MRE stripe. We demonstrated that these elements can create a system like a mechanical rotational spring. Additionally, it is easy to control the gripper by pulling the electromagnet. Compared to the ideas presented in \cite{Zhang20218181,Skfivan:8722762,Choi202044147,bernat2022gripper,Guan2022,bose2021magnetorheological,Cramer2018127} a new effect of the rolling torque between the MRE and a permanent magnet is exploited to provide a self-closing state. To verify the proposed concept, the gripper prototype was built and tested in experiments.

\section{The Gripper Concept}

The gripper presented in this work is based on magnetorheological elastomer and the permanent magnet. The first is elastomer material with soft magnetic particles. Its main features are flexibility and reactive to the electromagnet field. The second is a permanent magnet that stores a lot of magnetic field energy. In our work, we used these elements to create a flexible magnetic spring.

\subsection{The Magnetic Spring}

\begin{figure*}[htb]
    \centering
    \includegraphics[width=0.8\textwidth]{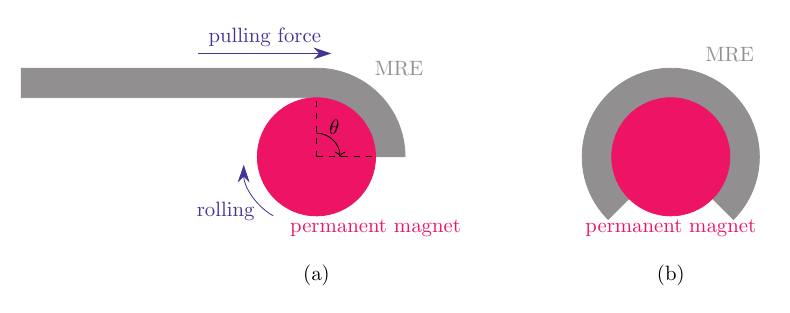}
    \caption{The working principle of MRE stripe attached to a permanent magnet. (a) pulling force causing rolling. (b) steady state.}
    \label{fig:workingPrinciple}
\end{figure*}

\begin{figure}[htbp]
    \centering
    \includegraphics[width=0.5\textwidth]{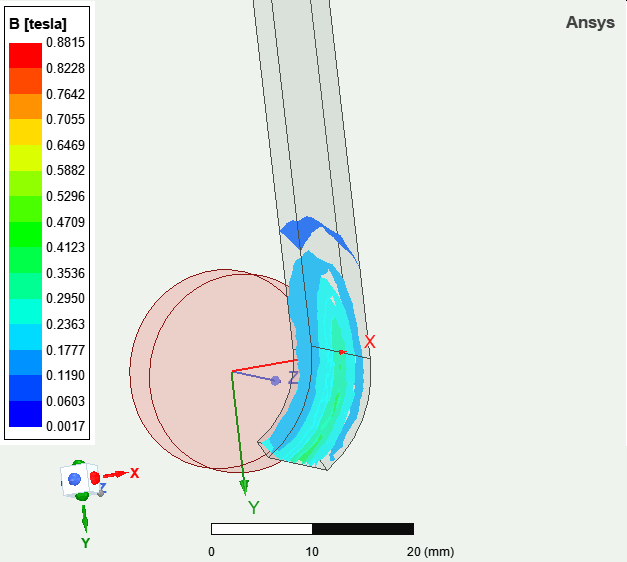}
    \caption{The magnetic field in MRE stripe attached to the permanent magnet with example contact angles equal to \SI{60}{\degree}.}
    \label{fig:magneticField}
\end{figure}

The main idea of the gripper is based on the magnetic spring created with a stripe built with MRE material and a cylindrical permanent magnet. Its behaviour can be illustrated in Fig. \ref{fig:workingPrinciple}. The magnetic field of the permanent magnet causes MRE to be screwed onto the permanent magnet. To show that the connection of the MRE stripe with the permanent magnet produces the rolling torque,  simulations were performed using ANSYS 2023 R1. The example of the field is shown in Fig. \ref{fig:magneticField}. It can be seen that the volume that contains the magnetic field is increasing with rolling MRE on the permanent magnet. Therefore, the torque is produced as long as MRE can be fully put on a permanent magnet. 

\begin{figure*}[htbp]
    \centering
    \subfloat[]{
        \includegraphics[width=0.5\textwidth]{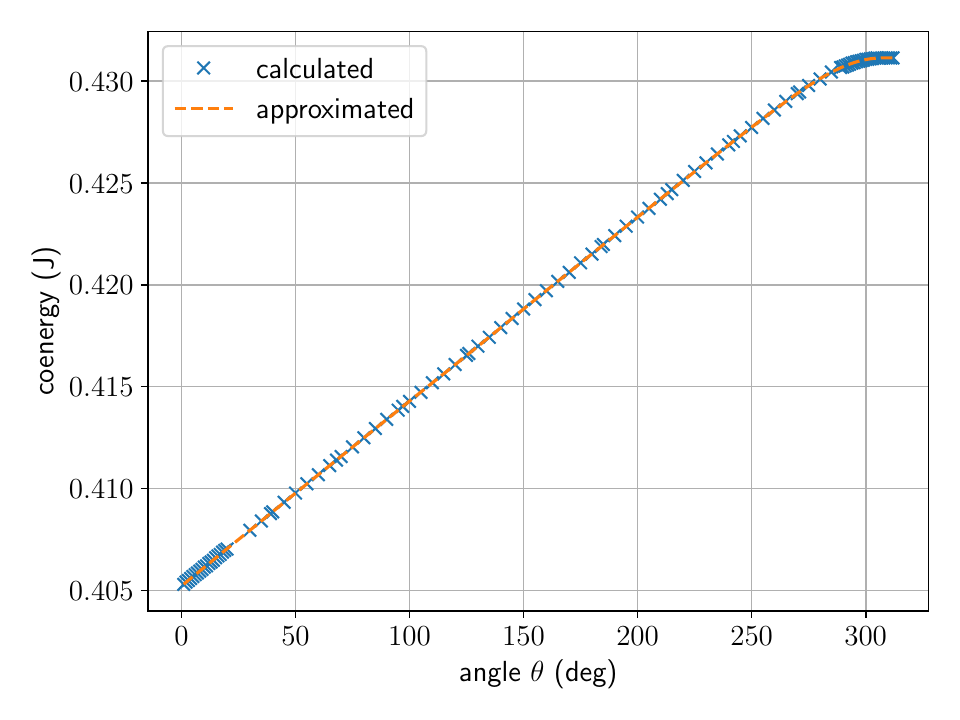}
    }
    \hfill
    \subfloat[]{
        \includegraphics[width=0.5\textwidth]{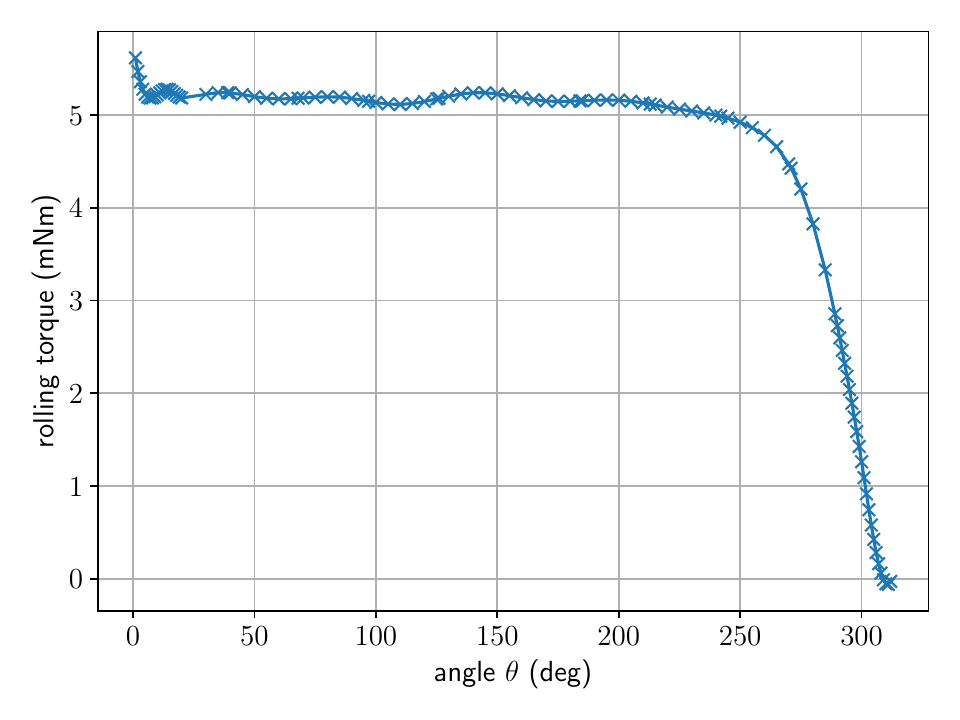}
    } 
    \caption{The magnetic coenergy versus angle for MRE stripe rolling on permanent magnet (a) and the force characteristics (b) calculated from virtual work principle.}
    \label{fig:rollingFEM}    
\end{figure*}

\begin{figure}[htb]
    \centering
    \includegraphics[width=0.4\textwidth]{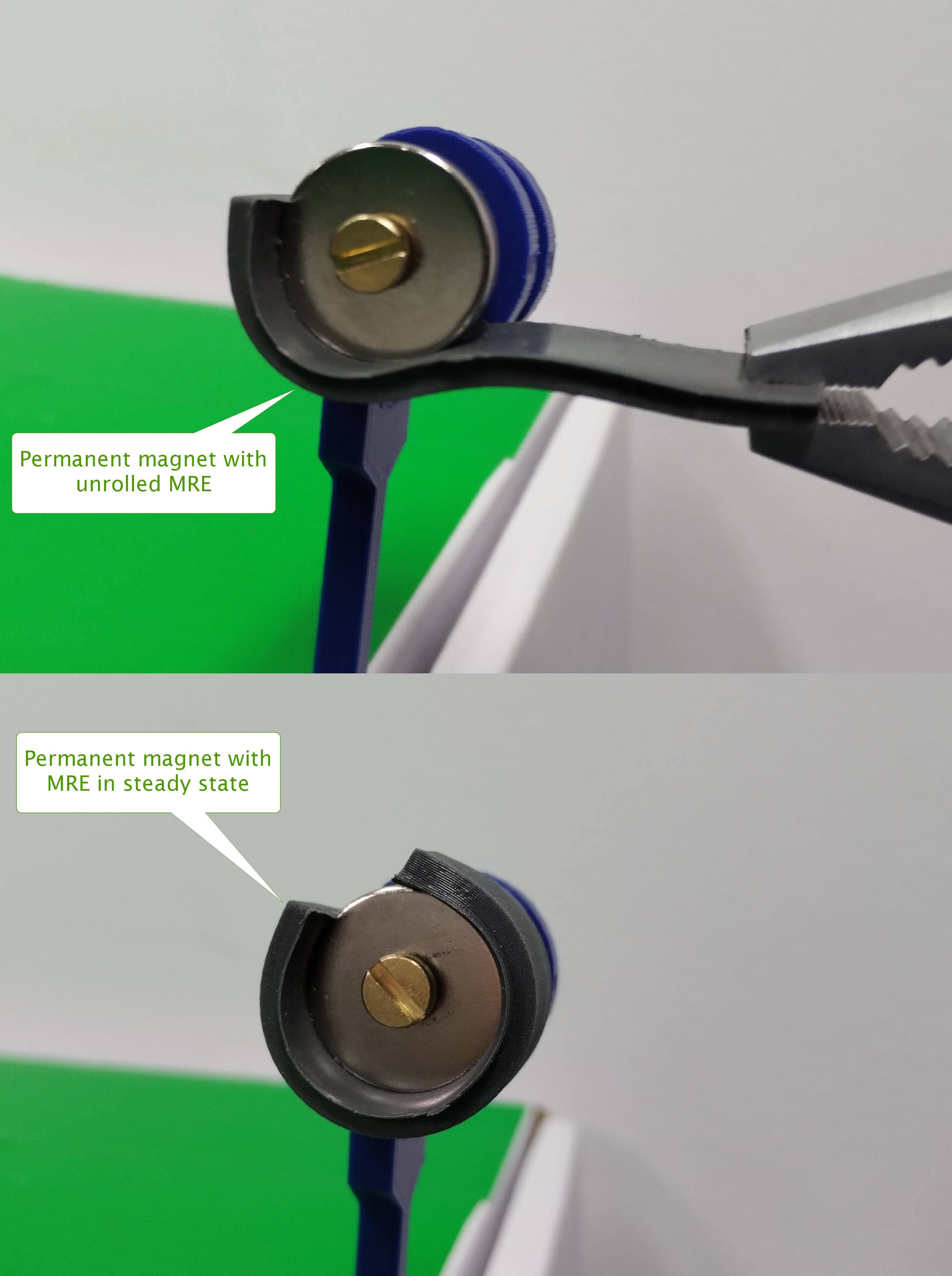}
    \caption{The cylindrical permanent magnet with MRE stripe rolled in.}
    \label{fig:rollingPM}
\end{figure}

The torque characteristic was calculated using the virtual work principle \cite{Ren19921212,Carpentier2014233}. The general expression for the coenergy is given as:
\begin{equation}
  W_{co}(\theta) = \int_{V}{\left(\int_{0}^{H}{BdH}\right)dV}
\end{equation}
where $\theta$ describes the angle between MRE and permanent magnet (presented in Fig. \ref{fig:workingPrinciple}), $V$ is the volume of the analyzed object, and $H$, $B$ are the magnetizing field and magnetic field respectively found by solving Maxwell's equation in the FEM software. The torque can be found by the derivative of the coenergy:
\begin{equation}
  T_{co}(\theta) = \frac{\partial W_{co}(\theta)}{\partial \theta}
\end{equation}
with respect to $\theta$. In our work, in the first step, the coenergy of MRE stripe and the permanent magnet was found for a range of $\theta$ angle using ANSYS software. Then, to eliminate numerical errors in solving Maxwell's equation, it was approximated by a spline function as shown in Fig. \ref{fig:rollingFEM}a. Finally, the torque $T_{co}(\theta)$ was found by calculating the derivative. The results are visible in Fig. \ref{fig:rollingFEM}b. It is clear that the rolling torque for the angle between \SIrange{0}{225}{\degree} is almost constant. If the MRE stripe is almost all rolled on the permanent magnet, then the torque is going to \SI{0}{mNm}. The behaviour can also be simply checked by experiment using the MRE stripe and a rolling permanent magnet, as shown in Fig. \ref{fig:rollingPM}.

\subsection{The Gripper Construction}

\begin{figure}[htb]
    \centering
    \includegraphics[width=0.4\textwidth]{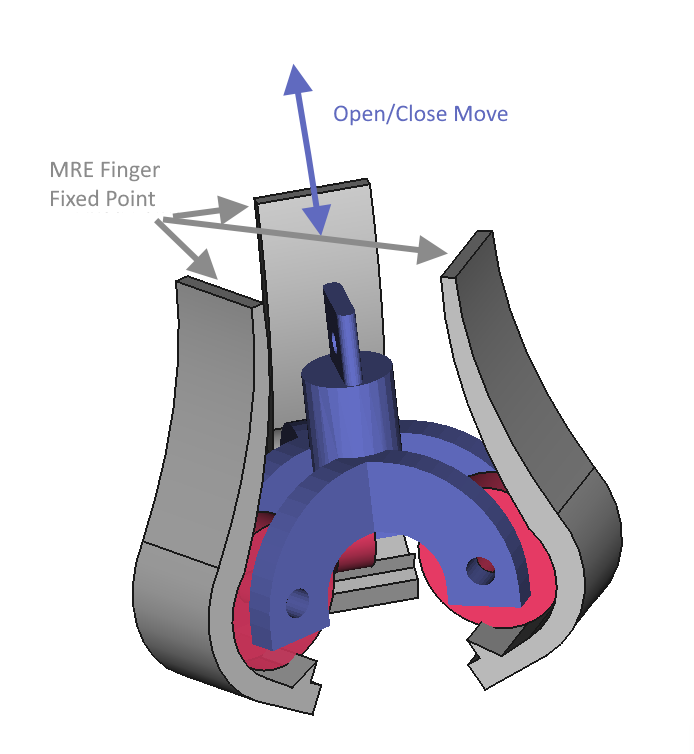}
    \caption{The idea of the gripper moves to the open and closed state. Elements: MRE (grey), permanent magnet (red), and mounting frame (blue).}
    \label{fig:concept}
\end{figure}

The gripper concept is to use three flexible springs as gripper fingers. The MRE stripes are fixed at the beginning and free to move on the permanent magnet. The field energy stored in a permanent magnet causes the MRE stripes to attempt to close. The MRE stripes have a small triangle near the handle object for a better grip. The concept is visualized in Fig. \ref{fig:concept}, where its open and closed states concept are presented. The beginning of the MRE stripes is fixed to some frame (not shown in the figure). In the open state, the extra force pulls the MRE stripes to make space for an object to be held. In the closed state, there is no extra force, and hence the permanent magnet and MRE elastomer (creating a flexible spring) are close to the element. The gripper shown in Fig. \ref{fig:concept} works by moving the mounting frame with permanent magnets. The moving frame has been designed in such a way that there was free space for holding objects between permanent magnets. However, the force can also be attracted to the MRE stripe, and hence the permanent magnet mounting frame has a fixed position. The advantage of MRE is the possibility to easily mount MRE in the frame and it interacts with a permanent magnet, giving the self-closing state.

\begin{figure}[htb]
    \centering
    \includegraphics[width=0.4\textwidth]{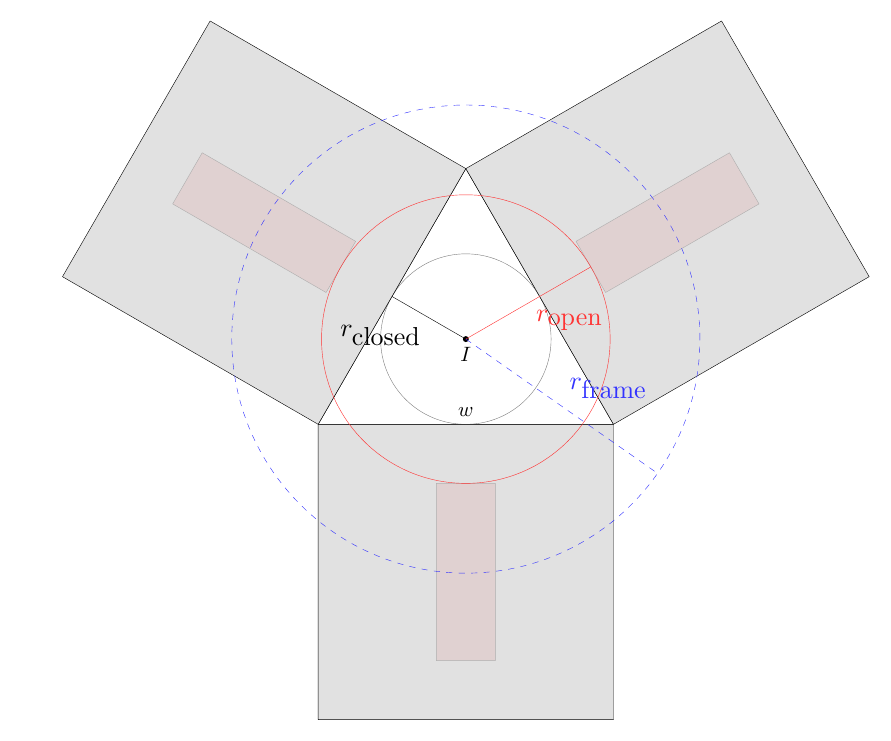}
    \caption{The bottom view of gripper fingers with marked cross-section space in open and closed state. Permanent magnets (red) and MRE fingers (gray).}
    \label{fig:grippingSize}
\end{figure}

The size of the gripped objects depends on the position of permanent magnets and MRE fingers. In Fig. \ref{fig:grippingSize} two marked circles are showing the space for cylindrical objects in the open and closed state. In the open state, objects are bounded by permanent magnets, which are shown by a circle with radius $r_{open}$. The following relationship $r_{open} = r_{frame} - \frac{1}{2}d_{PM}$ describes the connections between radius in the open state, mounting frame radius (place for the center of permanent magnets) $r_{frame}$ and permanent magnet diameter $d_{PM}$. In the closed state, objects are bounded by MRE fingers of minimal size given by a circle with radius $r_{close}$. It is easy to calculate using equilateral triangle relationships that the width of the finger is related to the radius in the closed state as $r_{close} = \frac{\sqrt{3}}{6}w$ where $w$ is the width of the MRE finger.

\subsection{Preparation of Magnetorheological Elastomers}

\begin{figure*}[htb]
    \centering
    \subfloat[]{
        \includegraphics[width=0.4\textwidth]{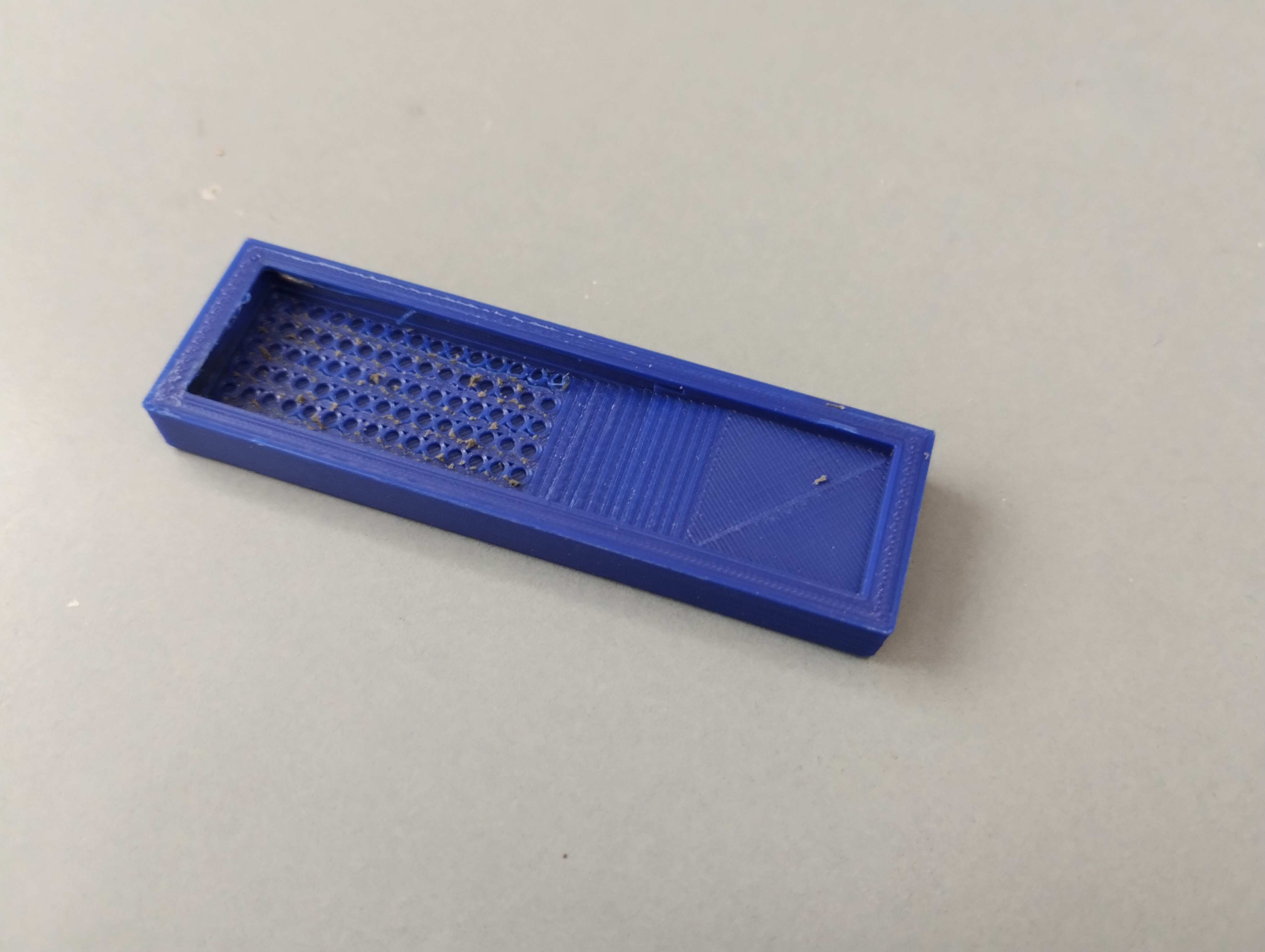}
        \label{fig:mold}
    }
    \hfill
    \subfloat[]{
        \includegraphics[width=0.4\textwidth]{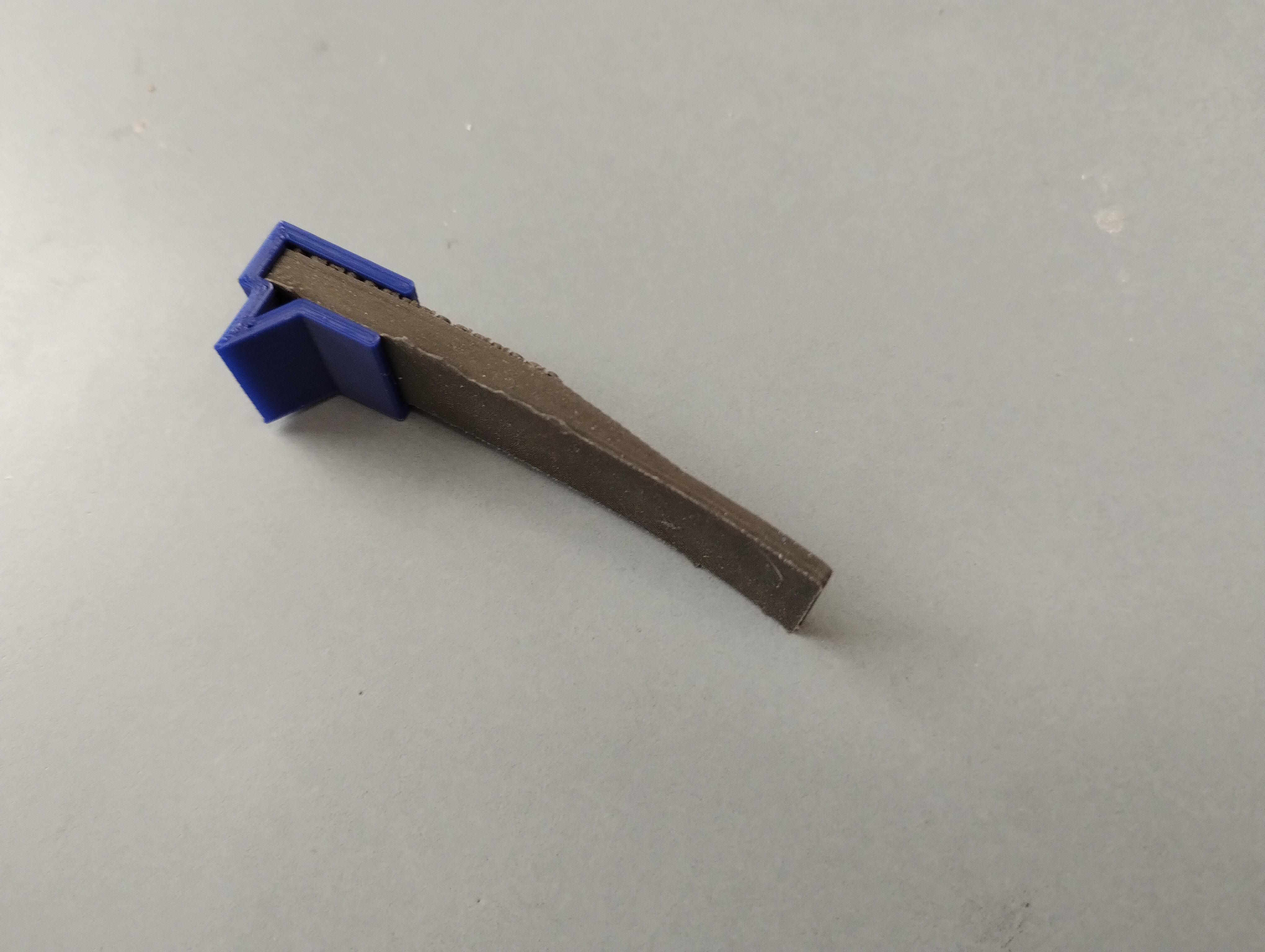}
        \label{fig:gripper_finger_with_model}
    }
    \hfill
    \subfloat[]{
        \includegraphics[width=0.4\textwidth]{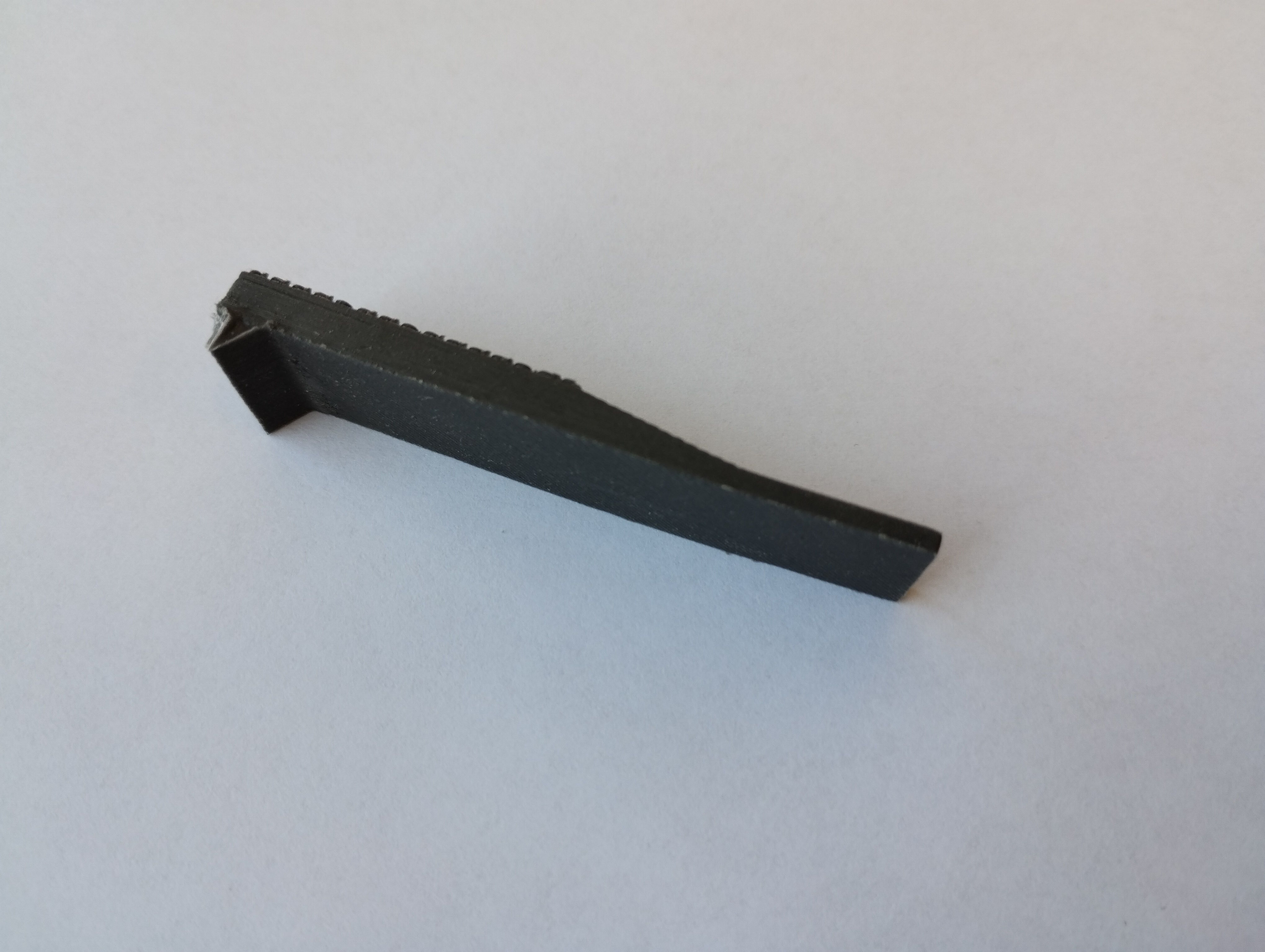}
        \label{fig:gripper_finger_alone}
    }
    \caption{Molds used during the preparation process of gripper finger (a), (b). The final gripper finger (c).}
\end{figure*}

The MRE material was prepared according to the procedure described in \cite{bernat2022gripper}. Three silicons with different properties were used to prepare MRE. These were Mold Star 15 (MS15) and Dragon Skin 10 (DS10) (both from Smooth-On) and RTV (OTT-S825 from OTTSilicone). The iron powder with an average particle size of \SI{63}{\micro\meter} was weighed in a glass container and thoroughly mixed with each silicone until a homogeneous mixture was obtained. The weight ratio of silicone to iron powder was equal to 1:1. Then, in the case of RTV silicone, the catalyst was added in an amount of 2\% by weight relative to the amount of silicone. The MS15 and DS10 are two-component silicons in which one of the components already contains a catalyst. After thorough mixing, the resulting mixture was degassed by vacuum treatment. Finally, the mixture was poured into the mold presented in Fig. \ref{fig:mold}. After 24 h, the MRE material was removed from the mold and carefully checked for bubbles in the structure. In the next step, the mixture of silicone with iron was prepared with the same procedure as before and the second part of the gripper was prepared (Fig. \ref{fig:gripper_finger_with_model}). Both parts of the mold were made by 3D printing. The final gripper finger is presented in Fig. \ref{fig:gripper_finger_alone}. 

\subsection{Mechanical properties of silicons and MREs}
The mechanical properties, i.e. tensile tests, of individual silicones and MREs made on their basis were examined to connect the properties of the obtained MREs with their action as actuators. Pure silicone samples (DS10, MS15, RTV) and MREs obtained from them were prepared as bars with dimensions of 50 mm length, 15 mm width, and 2 mm thickness. The tests were carried out at a traverse speed of 5 mm/min up to 30\% of sample strain for Young's modulus ($E_{mod}$) determination, and 50 mm/min for the remaining part of the test in order to determine the stress value at 100\% ($\sigma_{100\%}$) and 300\% ($\sigma_{300\%}$) sample strain. The initial force was \SI{0.05}{\newton}.

\section{The Experiments}

This section describes a series of experiments to verify the proposed concept. Firstly, the production of the gripper is discussed and then its validation is described.

\subsection{Gripper Construction}

\begin{figure*}[htb]
    \centering
    \subfloat[]{
        \includegraphics[width=0.476\textwidth]{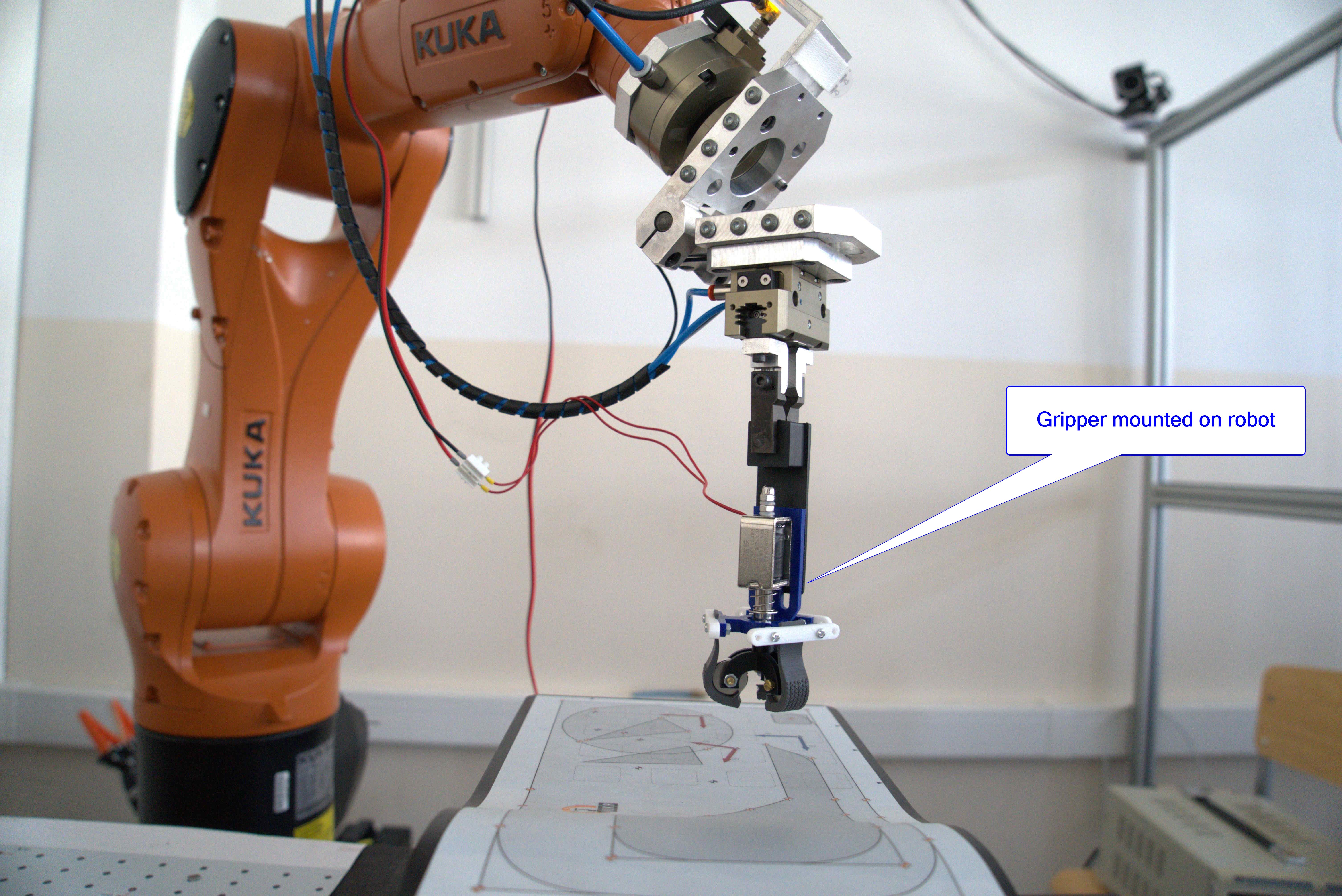}
        \label{fig:station}
    }
    \subfloat[]{
        \includegraphics[width=0.4\textwidth]{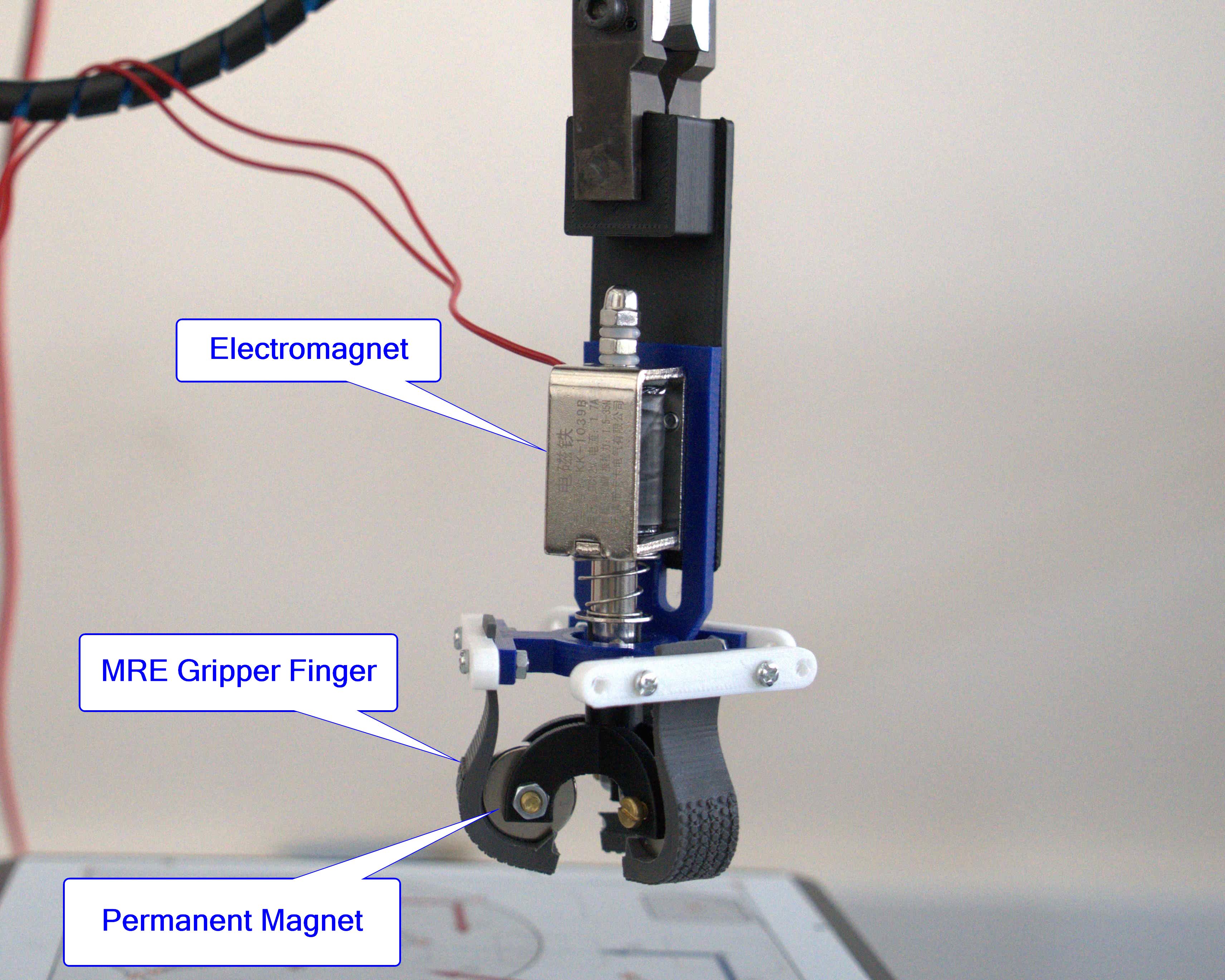}
        \label{fig:gripper}
    }
    \caption{The laboratory station with Kuka KR 6 R900 manipulator (a). Zoom of the gripper with mounted electromagnet (b).}
    \label{fig:gripperAndStation}
\end{figure*}

The gripper is constructed of three MRE stripes, three permanent magnets, a linear electromagnet, and a frame created by additive manufacturing. The parameters of the MRE finger, linear electromagnet, frame, and permanent magnet are provided in Table \ref{tab:dimensions}. Relying on them and the open/closed radius presented in Fig. \ref{fig:grippingSize} is equal to $r_{open}=$\SI{8}{\milli\meter} and $r_{close}=$\SI{4.33}{\milli\meter}. Movement of the frame with permanent magnets is ensured by a linear electromagnet with a spring defined in Table \ref{tab:dimensions}. In our gripper, the input signal is the voltage applied to the electromagnet. The turn on the gripper causes its opening and the turn off causes its closing. The concept of the gripper is presented in Fig. \ref{fig:gripper}, where the most important elements such as electromagnet (the element that allows controlling gripping process), permanent magnets (which allow rolling around them) and MREs (a crucial part of the gripper that allows gripping objects using its magnetic abilities and makes the gripper soft) are marked.

\begin{table}
\caption{\label{tab:dimensions}The MRE finger dimensions, electromagnet (EM), and neodymium permanent magnet (PM) properties (based on the manufacturer datasheet)}
\centering
\begin{tabular}{lll}
\toprule
Name (Symbol) & Value & Unit\\
\midrule
    MRE finger width ($w$) & 15 & \si{\milli\meter} \\
    MRE finger length & 60 & \si{\milli\meter} \\
    MRE finger thickness & 2-4 & \si{\milli\meter} \\
    Mounting frame radius ($r_{frame}$) & 18 & \si{\milli\meter} \\
    EM move range & 12 & \si{\milli\meter} \\
    EM rated current & 0.7 & \si{\ampere} \\
    EM rated voltage & 12 & \si{\volt} \\
    EM max load & 3.5 & \si{\kilo\gram} \\
    PM inner diameter & 4.2 & \si{\milli\meter} \\
    PM outer diameter ($d_{PM}$) & 20 & \si{\milli\meter} \\
    PM thickness & 5 & \si{\milli\meter} \\
    PM remanence ($B_r$) & 1.21-1.25 & \si{\tesla} \\
    PM coercivity ($H_c$) &  899 & \si{\kilo\ampere\per\meter} \\
\bottomrule
\end{tabular}
\end{table}

\subsection{Validation}

The aim of the validation process was to verify the basic principle of the gripper operation, its holding possibility of various objects, repeatedly holding process and measure the gripper finger pushing force.

\begin{figure}[htb]
    \centering
    \subfloat[]{
        \includegraphics[width=0.5\textwidth]{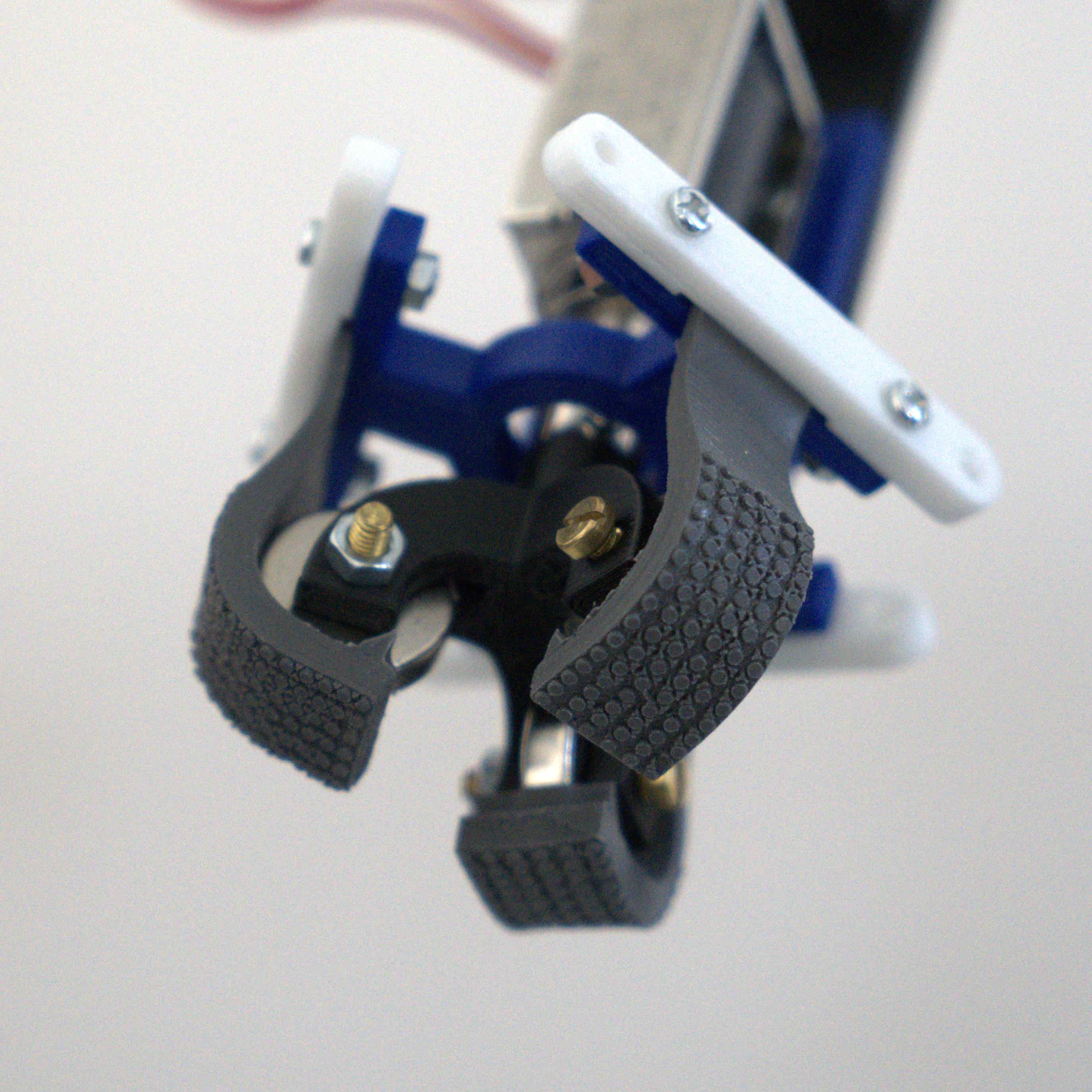}
        \label{fig:gripper-open}
    }
    \hfill
    \subfloat[]{
        \includegraphics[width=0.5\textwidth]{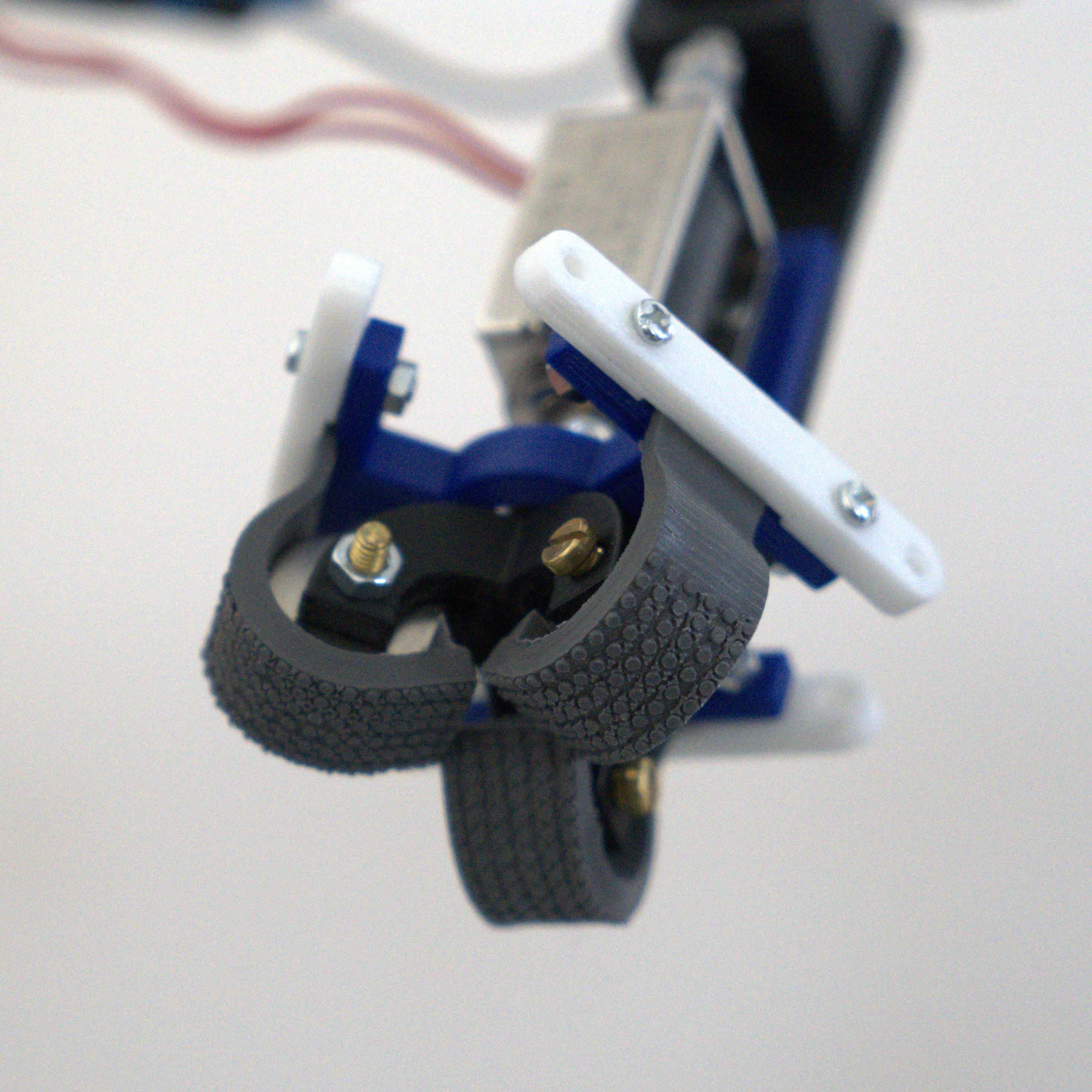}
        \label{fig:gripper-closed}
    }
    \caption{The gripper fingers in the open (a) and closed state (b).}
    \label{fig:basicValidation}
\end{figure}

The validation of the gripper was conducted by a series of experiments with the gripper mounted on the KUKA manipulator (KR 6 R900). Fig. \ref{fig:gripper-open} and \ref{fig:gripper-closed} show two states of this device: a) this is the first situation when the gripper is open, whereas b) shows what happens when the gripper is closed and the rolling effect around the permanent magnets is visible. In Fig. \ref{fig:gripper-closed} it can be seen that the rolling around of the MRE stripes causes them to come closer to each other and at this point, the grip occurs. According to Fig. \ref{fig:grippingSize} the radius $r_{closed}$ can be slightly lower due to the possibility of compressing MRE finger corners when touching. 

\begin{figure}[htbp]
    \centering
    \subfloat[]{
        \includegraphics[width=0.5\textwidth]{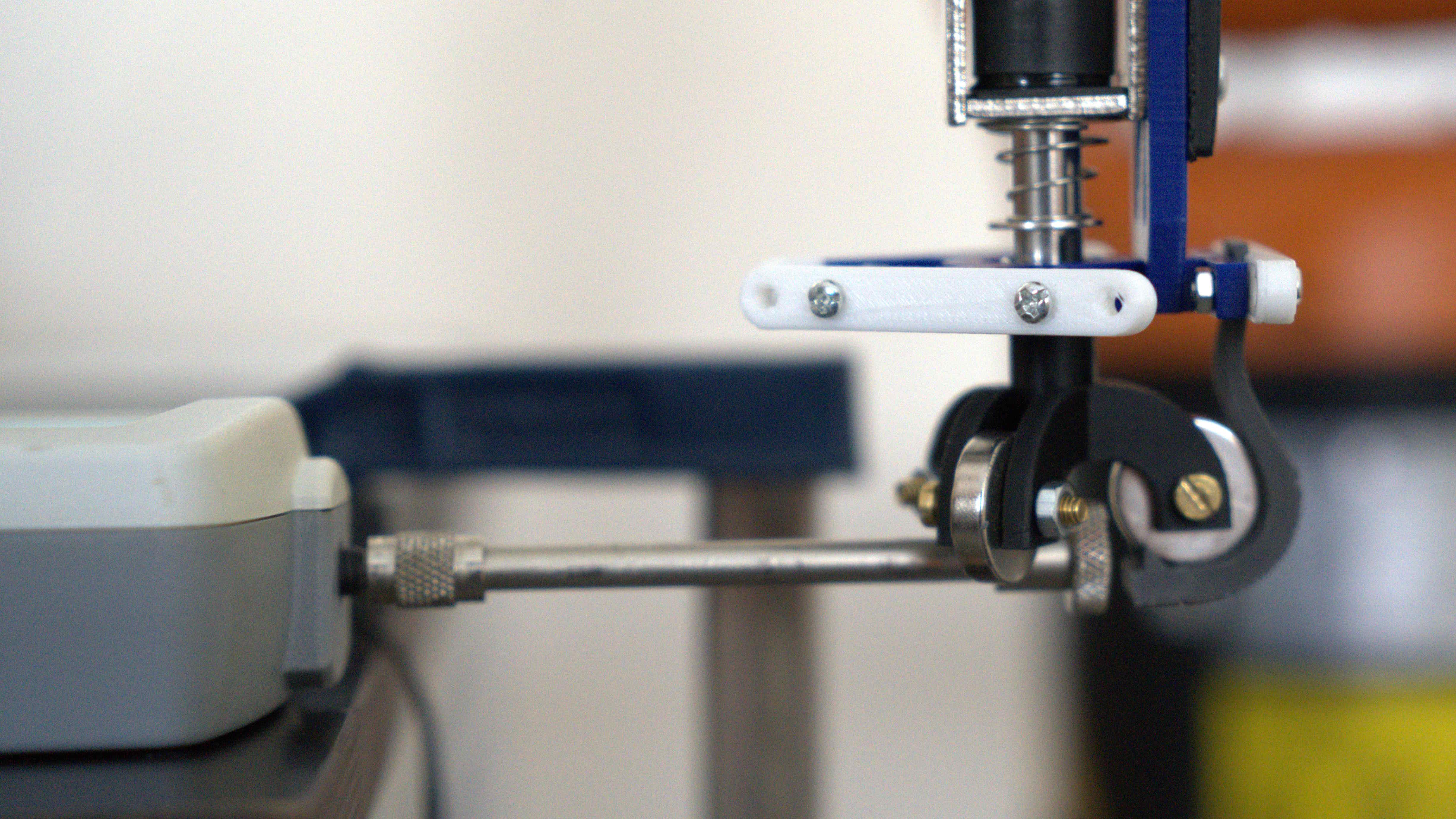}    
    }
    \\
    \subfloat[]{
        \includegraphics[width=0.5\textwidth]{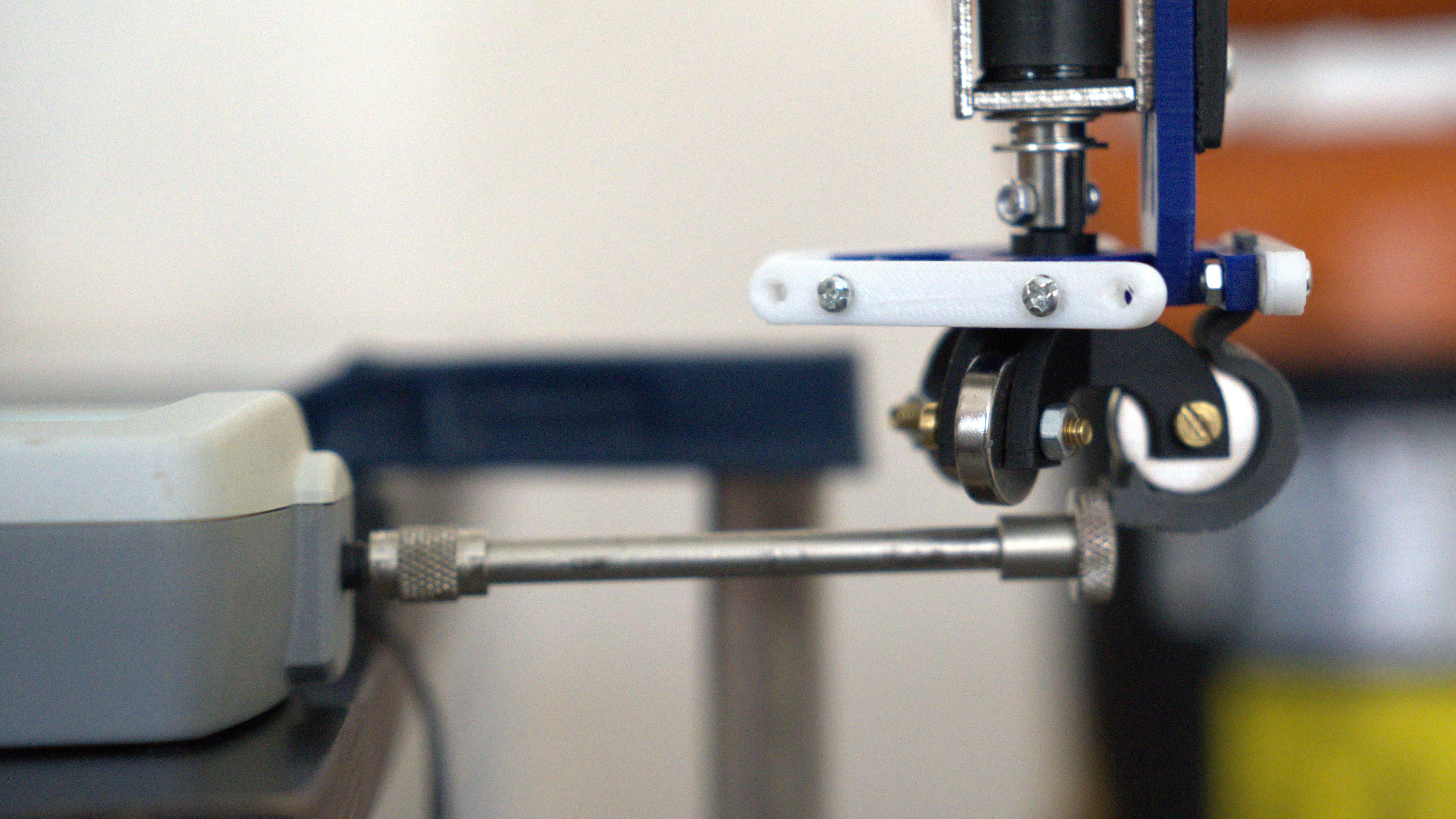}    
    }
    \caption{The gripper finger pushing the force sensor in the closed state. Open state (a) and closed state (b).}
    \label{fig:force-sensor}
\end{figure}

\begin{figure}[htbp]
    \centering
    \includegraphics[width=0.5\textwidth]{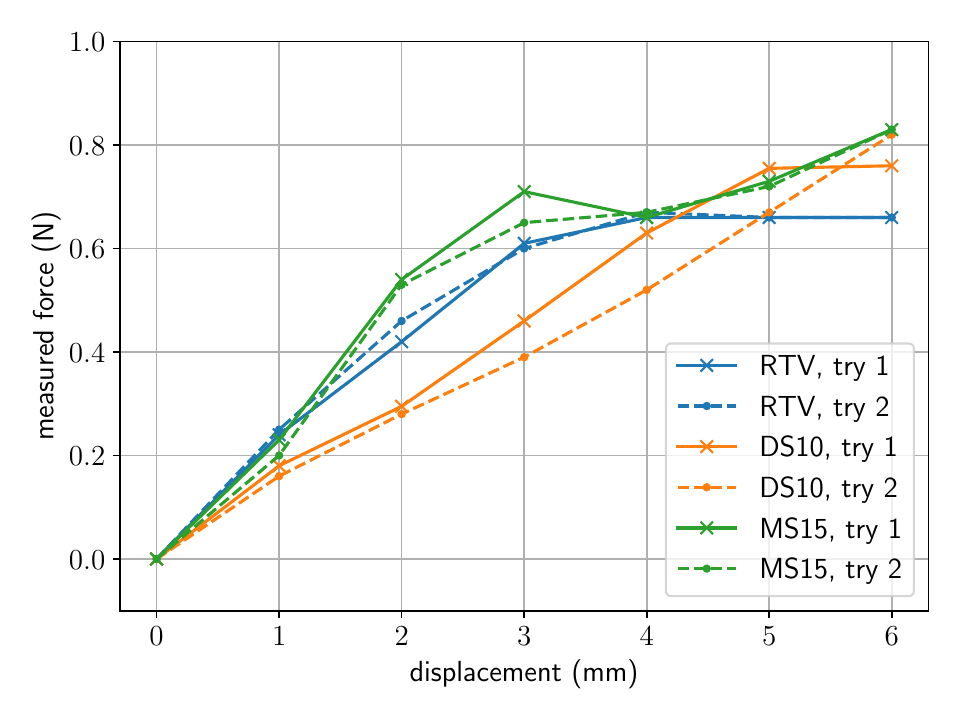}
    \caption{The gripper finger pushing force characteristics versus displacement.}
    \label{fig:forceDisplacement}
\end{figure}

An additional experiment was conducted on the single gripper finger to get to know the force acting in the closed state. The gripper with removed two fingers was set up with the force sensor (Axis FB20) as presented in Fig. \ref{fig:force-sensor}. The gripper was turned on/off five times. Each time, in the closed gripper state the force was read from the sensors. The average force is \SI{0.7\pm0.02}{\newton} and it shows that gripper fingers produce force when they are rotated on permanent magnets. In the next experiment, the finger was in the closed state, and the force sensor was set up almost touching. Then, the robot was moved with step \SI{1}{\milli\meter} touching the force sensor. The resulting characteristics are visible in Fig. \ref{fig:forceDisplacement}. It is visible that all materials have almost the same characteristics.

The experiment showing the operation of the gripper, its maximum range of grasp, and lifting capacity were carried out on 6 different objects. Tab. \ref{tab:elements} presents the specification of lifting elements (in the case of irregular shape, diameter describes the outer outline). All elements were successfully gripped 10 times for 10 tries (only the bilberry has an 8/10 ratio). The overall success rate of gripping is equal to \SI{96.6}{\percent}. The maximum mass of the lifted object was \SI{28}{\gram} (brass element). The speed of closing or opening of the gripper was estimated based on the movie. The closing time is about \SI{150}{\milli\second} and opening time is about \SI{175}{\milli\second}. 

\begin{table}
\caption{\label{tab:elements}Lifted elements with their mass and dimensions (W-width, L-length, H-height, D-diameter).}
\centering
\begin{tabular}{lll}
\toprule
    Name (Figure) & Mass & Dimension\\
\midrule
    Grape (\ref{fig:handle}a) & 2.50g & H=18\si{mm}, D=15mm\\
    Bilberry (\ref{fig:handle}b) & 2.00g & H=13mm, D=16.5mm\\
    Candy (\ref{fig:handle}c) & 16.34g & H=40mm, W=25mm, L=17mm\\
    Silicon hat (\ref{fig:handle}d) & 2.75g & W=25mm, H=28 mm\\
    Brass element (\ref{fig:handle}e) &  28.00g & H=25mm, D=22mm\\
    White ball (\ref{fig:handle}f) &  2.07g & D=20mm \\
\bottomrule
\end{tabular}
\end{table}

\begin{figure*}[htbp]
    \centering
    \subfloat[]{
        \includegraphics[width=0.3\textwidth]{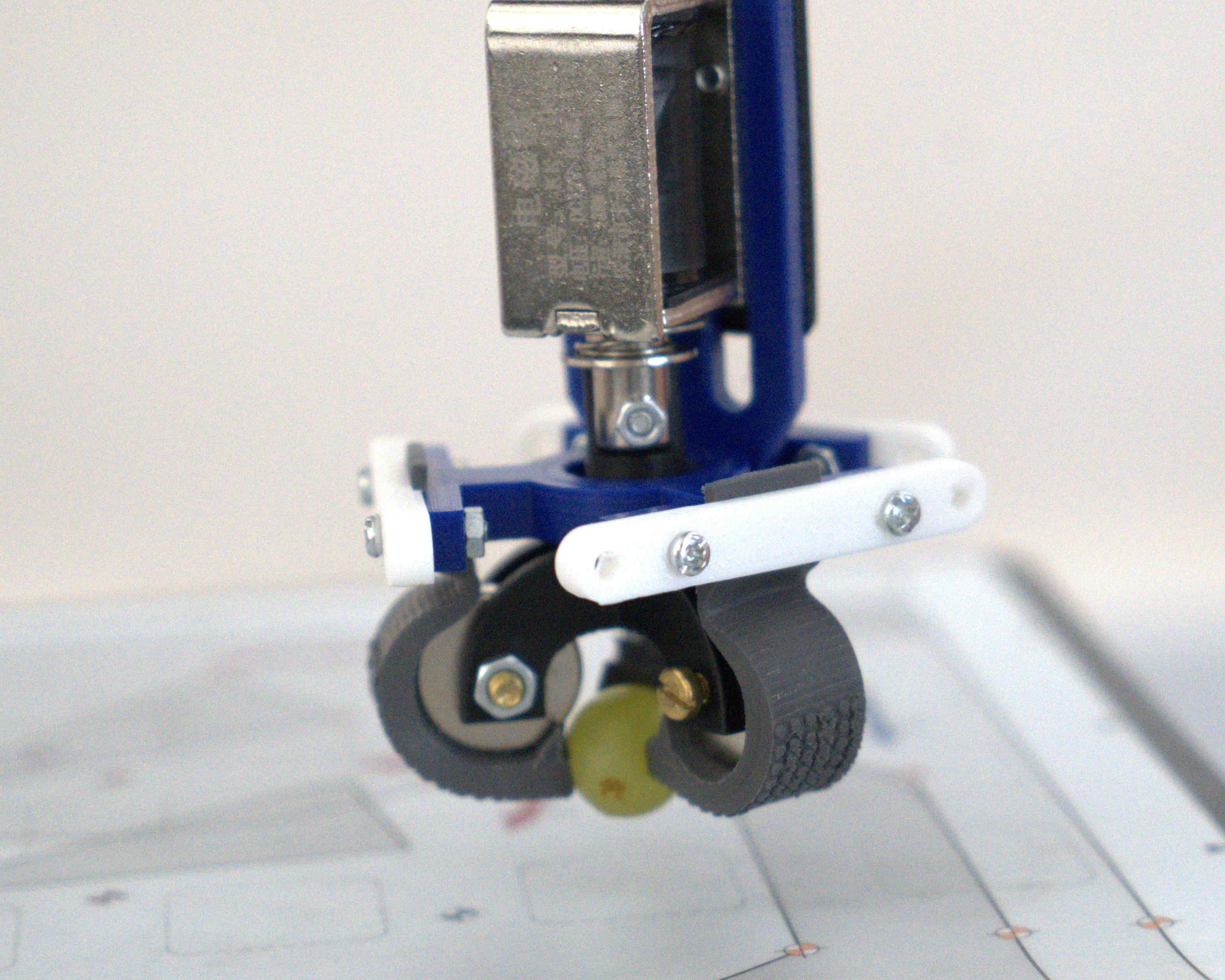}
    }
    \hfill
    \subfloat[]{
        \includegraphics[width=0.3\textwidth]{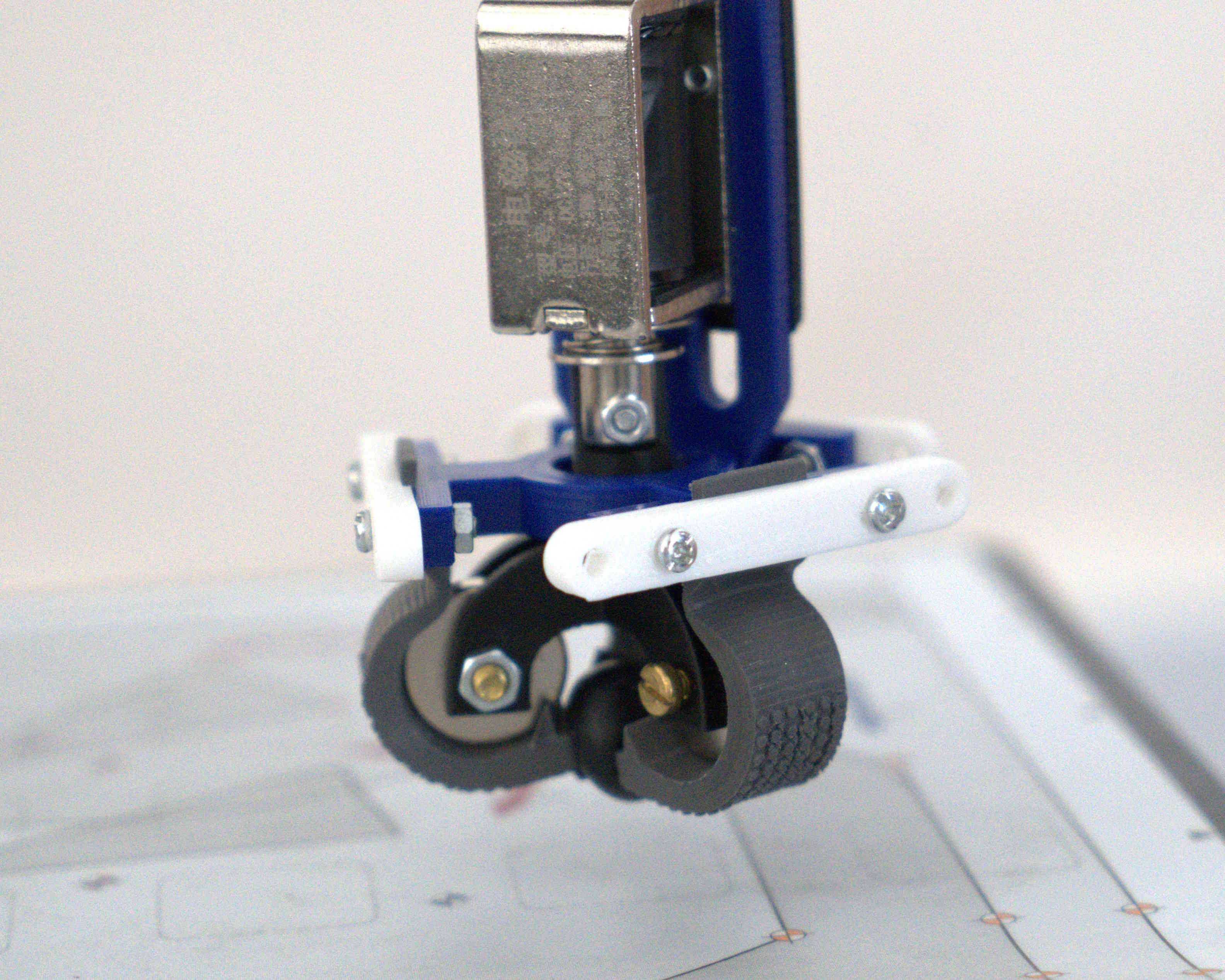}
    }
    \hfill
    \subfloat[]{
        \includegraphics[width=0.3\textwidth]{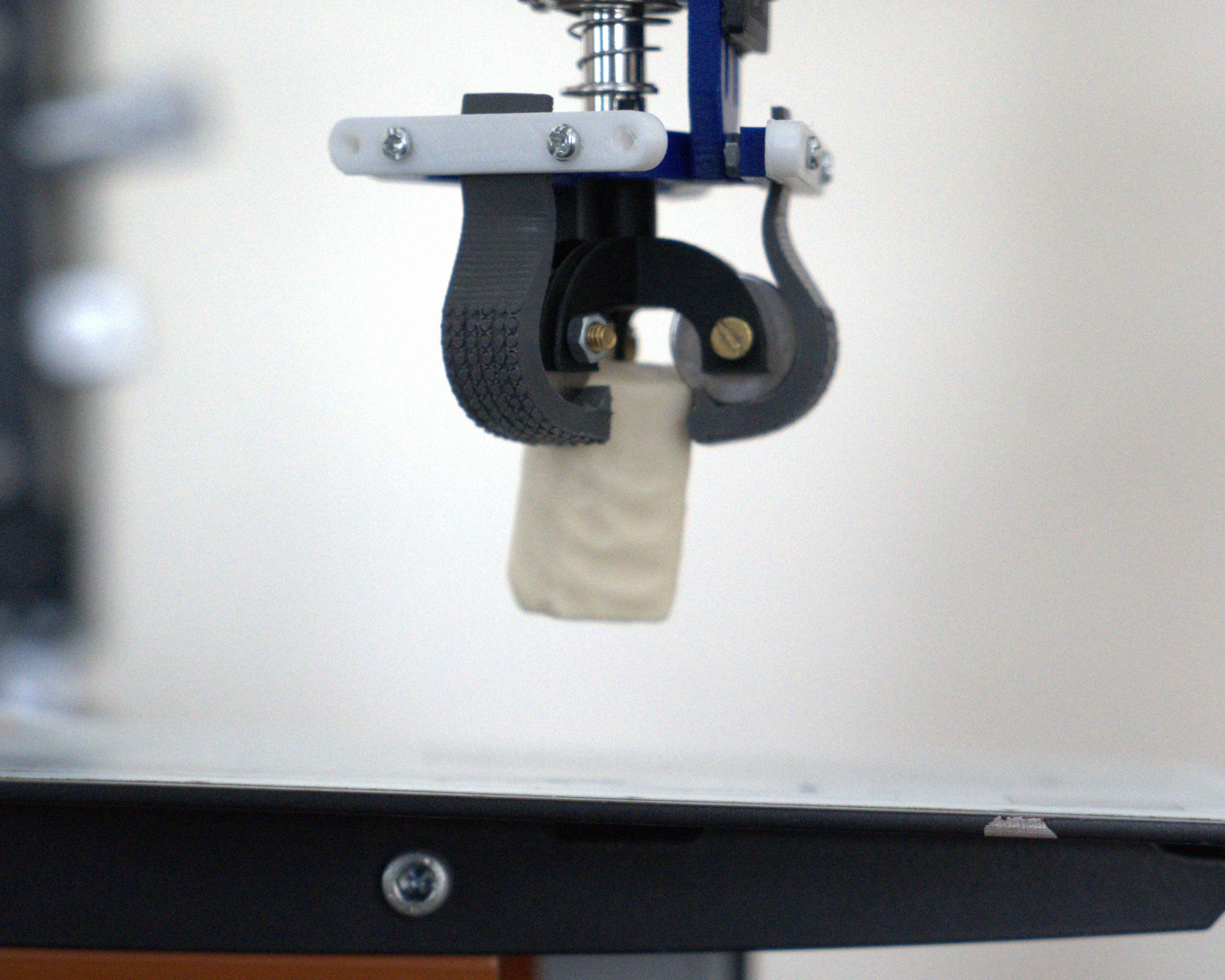}
    }
    \\
     \subfloat[]{
        \includegraphics[width=0.3\textwidth]{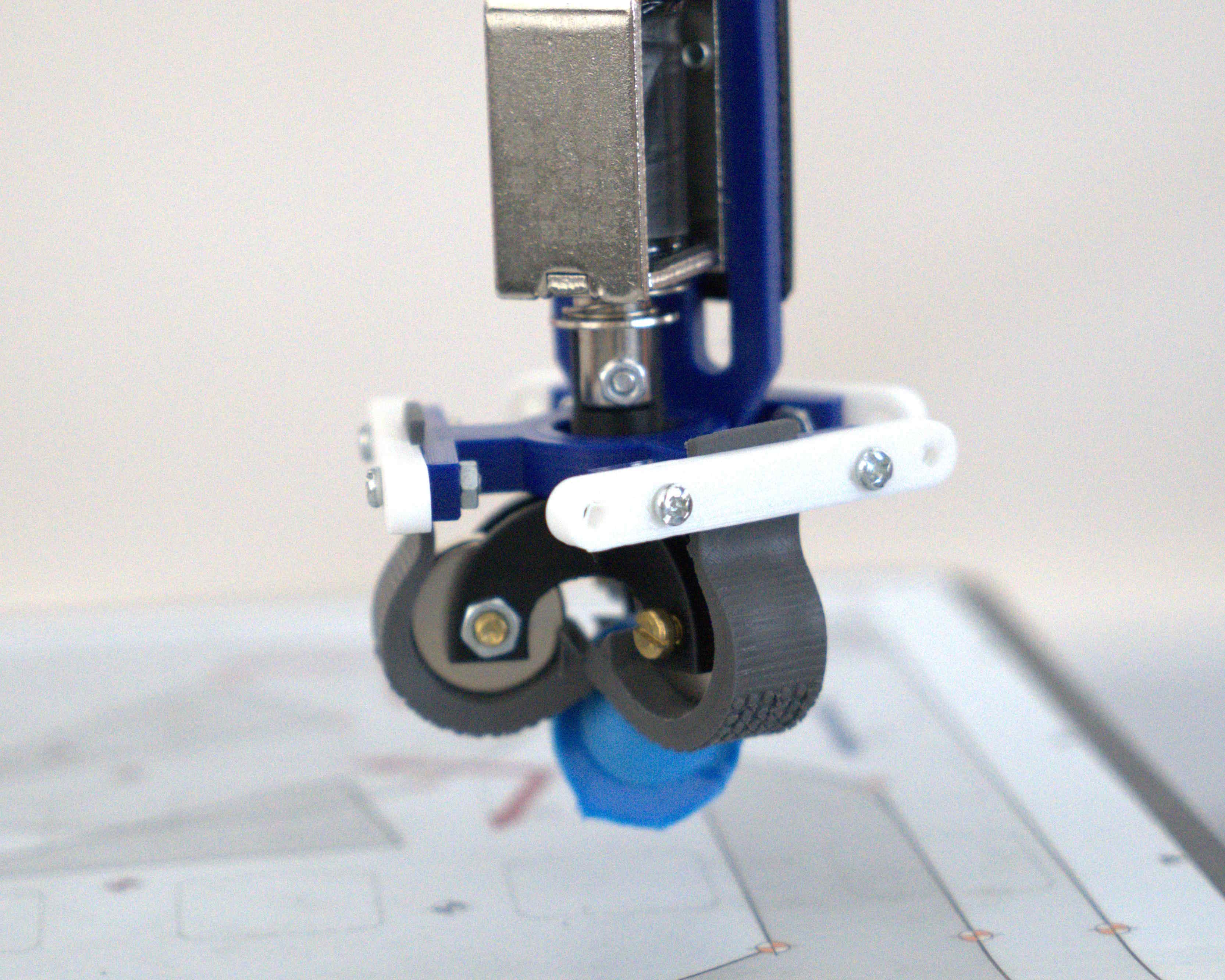}
    }
    \hfill
    \subfloat[]{
        \includegraphics[width=0.3\textwidth]{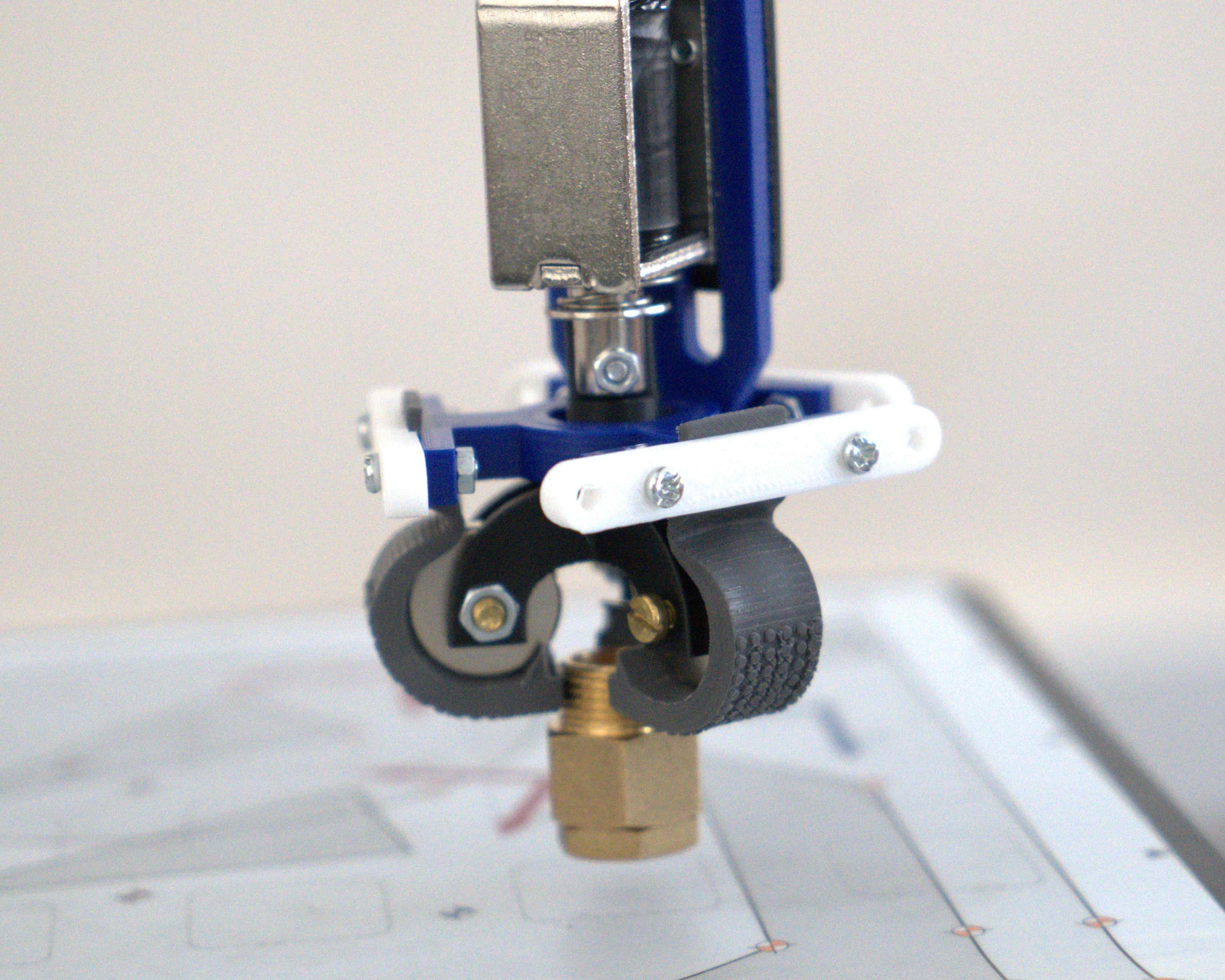}
    }
    \hfill
    \subfloat[]{
        \includegraphics[width=0.3\textwidth]{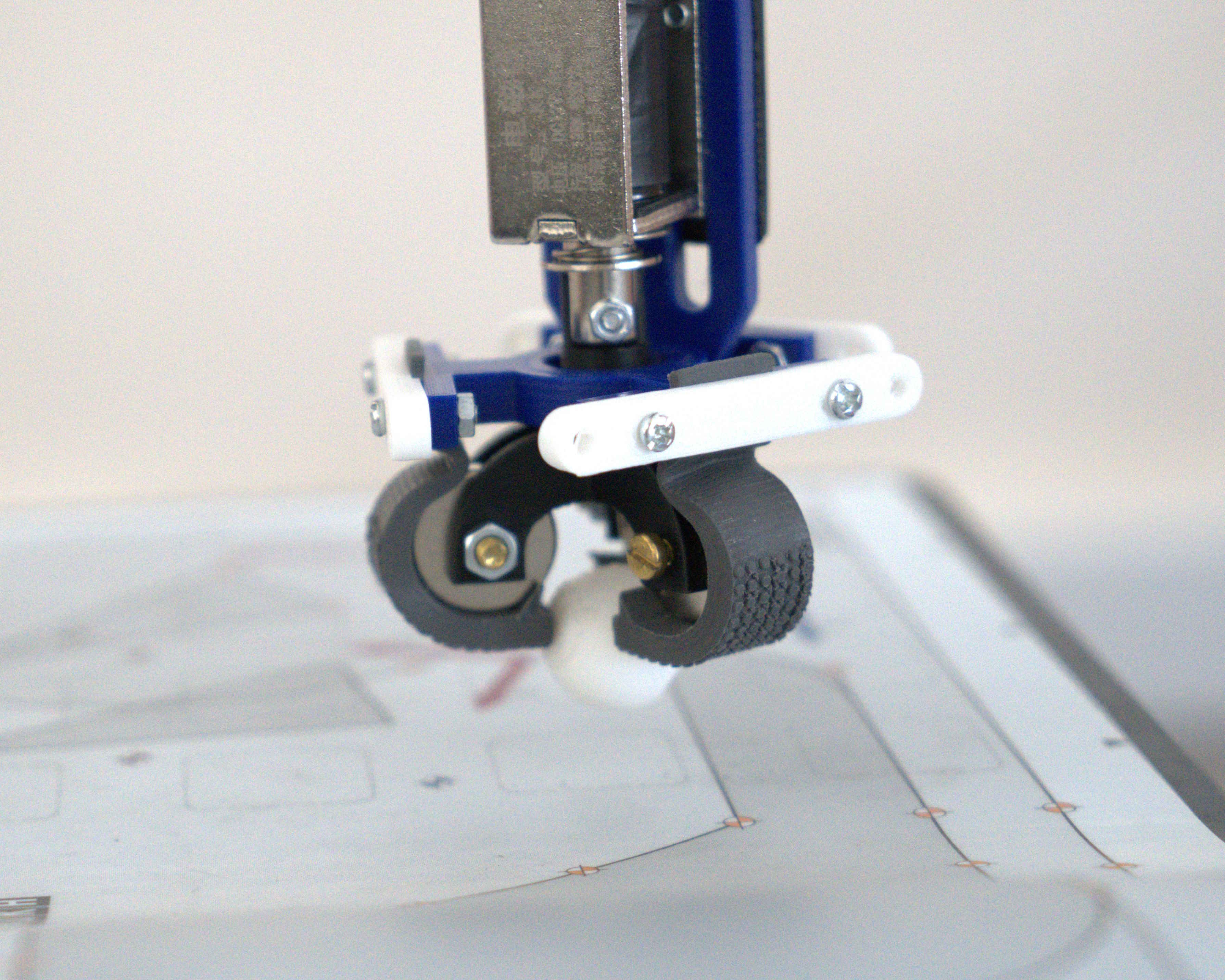}
    }
    \caption{The proposed MRE gripper handling various objects: grape (a), bilberry (b), candy(c), silicon hat (d), brass element (e), and white ball (f).}
    \label{fig:handle}
\end{figure*}

What is very important all photos in Fig. \ref{fig:handle} were taken in the same camera position and gravity acts vertically downwards. As presented, the orientation of the gripper does not affect gripping ability (so it is independent of gravity). The properties of MRE, its rough surface, and the strong magnetic force result in a firm gripping effect, which means that all types of elements used for validation did not slip out.

\begin{figure}[htb]
    \centering
    \includegraphics[width=0.5\textwidth]{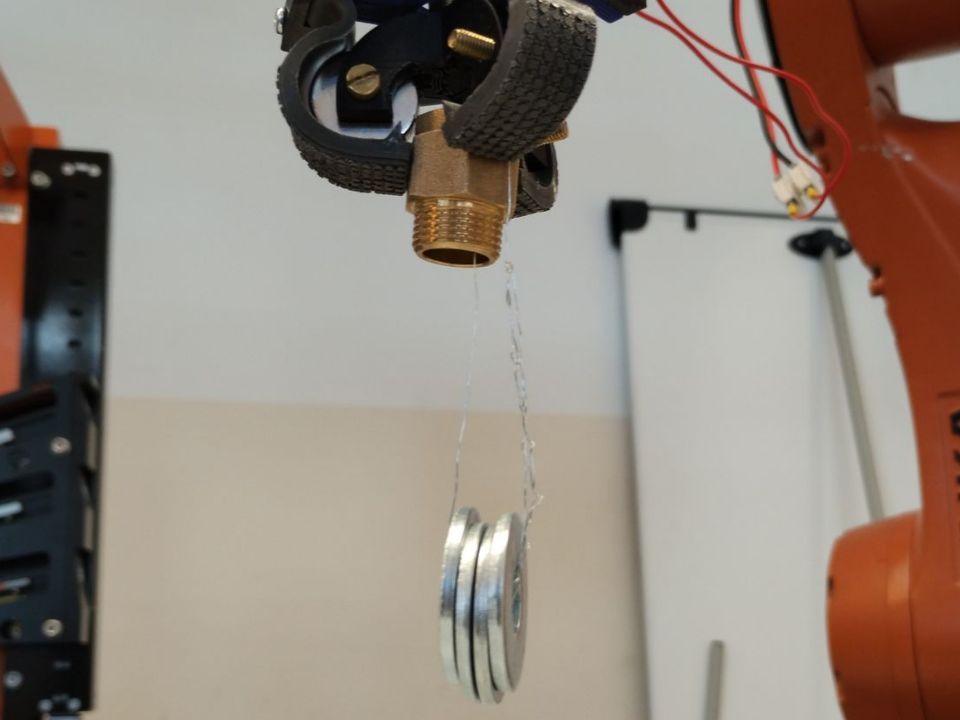}
    \caption{The experimental measurement of the gripping force with varying load}
    \label{fig:maxMass}
\end{figure}

Finally, the test of all materials was performed to find the maximum mass that the gripper could hold. The tests were performed with brass elements with attached additional metal washers as it is visible in Fig. \ref{fig:maxMass}. Table \ref{tab:maxMass} shows that the MRE prepared with RTV silicone can lift the largest mass, while the MRE prepared with DS15 silicone can lift the smallest mass, which is 2.4 times smaller than in the case of MRE with RTV silicone. The obtained results can be related to the mechanical properties, more precisely the tensile strength, of the obtained MREs. With the increase of stiffness (higher $E_{mod}$ values) and strength of materials (higher values of $\sigma_{100\%}$ and $\sigma_{300\%}$), the maximum mass that can be lifted by the gripper made of MREs increases.

\begin{table}
\caption{\label{tab:maxMass}The mechanical properties of pure silicones and MREs for RTV, MS10, and DS15 silicones. In the last column, the maximum mass lifted by the gripper with MRE fingers.}
\centering
\begin{tabular}{lllll}
\toprule
Name & $E_{mod}$, \si{\mega\pascal} & $\sigma_{100\%}$, \si{\mega\pascal} & $\sigma_{300\%}$, \si{\mega\pascal} & Maximum Mass, \si{\gram} \\
\midrule
    pure RTV  & 0.52 & 0.87 &   -  & - \\ 
    pure MS10 & 0.45 & 0.72 & 4.54 & - \\ 
    pure DS15 & 0.22 & 0.28 & 2.13 & - \\ \midrule 
    MRE with RTV  & 0.81 & 1.64 &   -  & 97.4\\ 
    MRE with MS10 & 0.78 & 0.93 & 4.04 & 86.5\\ 
    MRE with DS15 & 0.36 & 0.48 & 2.24 & 40.2\\ 
\bottomrule
\end{tabular}
\end{table}

\subsection{Discussion}

In the presented gripper the object grasping is based on the interaction between permanent magnet and magnetorheological elastomer stripe. The combination of rolling magnetorheological elastomer on the permanent magnet for creating the gripper is a novel concept from the point of view of author knowledge in literatures \cite{Zhang20218181,Skfivan:8722762,Choi202044147,bernat2022gripper,Guan2022,bose2021magnetorheological,Cramer2018127}. 

In general, the comparison of soft grippers quantitative indicators like lifting mass is difficult due to its variety of geometry and size - all grippers presented in works \cite{Zhang20218181,Skfivan:8722762,Choi202044147,bernat2022gripper,Guan2022,Choi2018,Choi2023} have different properties. In most cases, the same gripper design can be scaled to be larger or to be smaller to obtain different indicators.

In this work, the presented construction can create a normally open or closed state and hence it does not require power to hold objects or to be opened. Furthermore, from the author's point of view, the gripper is easy to produce with commonly accessible materials. The gripper is controlled by voltage excitation which is easily accessible and has fast response time. It has a much bigger maximum load capability (up to 10 times greater) in comparison to the finger based grippers defined in work \cite{bernat2022gripper,Guan2022}. Alternatively, the soft gripper based on the suction cup \cite{Zhang20218181} has a lifting ability of about \SI{300}{\gram} with a current of about \SI{10}{\ampere}, but this kind of gripper cannot manipulate objects (has no fingers). In our case, we energised the electromagnet \SI{12}{V} and \SI{0.7}{A}, which makes our gripper more energy efficient.

Furthermore, in our solution, the behaviour of the gripper fingers does not depend on the direction of gravity. Relying on the experiments, it was visible that the gripper process is very stable. This means that the gripper catches the object on the first try in most of the experiments. Furthermore, the speed of the gripper is very fast in comparison to work \cite{Skfivan:8722762} (up to 4 times).

In further work, it is possible to independently control each gripper finger. This gives the ability to manipulate objects in a more sophisticated way than just grabbing. Furthermore, the authors believe that maximisation of the gripping force by providing anisotropy to MRE material with shape optimisation is possible because the study on the anisotropy in MRE materials shown improvement as summarised in work \cite{bose2021magnetorheological}. 


\section{Conclusion}
In summary, the presented work introduces magnetorheological elastomer rolling on the permanent magnet, which is a novel mechanism in contrast to previous soft magnetic robots presented in the literature. The proposed mechanism is applied to construct a novel soft gripper. The main advantages of the proposed soft gripper are as follows. It is simply to build with readily available materials. As shown in the experiments, it can hold a variety of items, such as 3D-printed elements, a ball, a silicon cap, or brass elements. The overall success rate of gripping is almost \SI{97}{\percent}. The gripper can hold an element weighing up to about 97 grams, which is enough to lift small everyday usage objects like fruits or candies. Furthermore, the usage of soft fingers enables one to apply to grasping delicate objects.

\section*{Acknowledge} 
This research was funded by the Ministry of Education and Science, grant number 0211/SBAD/0123. The authors would like to thank dr Paweł Szulczyński for help in conducting of experiments with a KUKA robot.

\bibliographystyle{unsrtnat}
\bibliography{references}  






\end{document}